\documentclass[prl,aps,twocolumn]{revtex4}

\def\<{\langle} \def\>{\rangle}

\newcommand{\be}{\begin{equation}}

\newcommand{\bea}{\begin{eqnarray}}
\newcommand{\eea}{\end{eqnarray}}

\usepackage{graphics}
\usepackage[]{color}

\begin{document}

\title{Secure communication using mesoscopic coherent states}
\author{Geraldo A. Barbosa$^*$, Eric Corndorf, Prem Kumar, and Horace P. Yuen}
\address{Center for Photonic Communication and Computing, \\Department of Electrical and Computer Engineering,
Northwestern University, Evanston, IL  60208}
\date{Revised version: 04/17/03.}


\begin{abstract}
We demonstrate theoretically and experimentally that secure
communication using intermediate-energy (mesoscopic) coherent
states is possible. Our scheme is different from previous quantum
cryptographic schemes in that a short secret key is explicitly
used and in which quantum noise hides both the bit and the key.
This encryption scheme can be optically amplified.  New avenues
are open to secure communications at high speeds in fiber-optic or
free-space channels.
\end{abstract}

\maketitle

\vspace{4mm}


For the encryption of data with perfect secrecy \cite{shannon}
that cannot be broken with any advance in technology, one may in
principle employ one-time pad with secret key obtained by the BB84
\cite{bennett-brassard} quantum cryptographic technique for key
expansion. Such an approach is possible \cite{alamos}, however, it
is slow and inefficient because the key length needs to be as long
as the data, and it also requires a nearly ideal quantum
communication line that is difficult to obtain in long distance
commercial systems such as the Internet core. On the other hand,
for both military and commercial applications, there are great
demands for secret communications that are fast and secure but not
necessarily perfectly secure. (There are many practical issues,
human as well machine based, that would make theoretical perfect
security in specific models not so important in real life
\cite{schneier}). In the following, a new scheme based on ideas
similar to those of Ref. \cite{yuenAKE} is described for secure
{\em data encryption} that can be operated at optical speeds with
conventional optical technology, and a prototype experimental
implementation is presented. In this scheme, a short secret key is
classically extended and then used to encrypt data  in a way that
the quantum noise of the coherent states protects both the data
and the key.

The following line of reasoning describes the ideas
\cite{yuen_tobe} that led to the development  of our present kind
of quantum cryptographic schemes. One crucial element for
obtaining security in BB84 involves the detection of small
intrusion on weak signals, which is difficult to achieve in a
network environment.
 This problem would be alleviated if quantum signal sets of higher
 energy
are selected for different bit values by a secret key shared between the sender (Alice) and the receiver (Bob).
It is important to remember that some shared secret key is needed in BB84 for
message authentication during protocol execution. The resulting scheme is acceptable as  key expansion
 if the new key is secure even if the shared secret key is known to the
attacker after the user communications are completed.
When a secret key is used to identify the signal
set, it would be a secret CDMA (Code Division Multiple Access) scheme classically, which does not
allow key expansion because the user and the attacker have the same observation. We would discuss
elsewhere how a corresponding KCQ (Keyed CDMA in Quantum Noise) scheme can be used to obtain key
expansion in the quantum case. In this paper, we are concerned with the use of KCQ for data encryption.

There are two basic problems with classical encryption that does not employ the inefficient one-time pads.
 One is that the total data uncertainty $H({\bf X})$ given observation ${\bf Y}$ is bounded above by the key uncertainty,
$H({\bf X})\leq H({\bf K})$ \cite{shannon}. The other is that the key ${\bf K}$ may be found by a {\em known-plaintext attack}  when the eavesdropper (Eve) knows
 the output-input pairs $({\bf Y},{\bf X})$ for some data length. In our scheme,  $H({\bf X})$ is not bounded by $H({\bf K})$
because Eve cannot have the observation ${\bf Y}$ that Bob obtained via the {\em optimal
quantum} measurement utilizing the key ${\bf K}$. To extract information from even a full copy of the quantum signal
without knowing ${\bf K}$, Eve has to make a sub-optimal measurement that would yield information on all possible signal sets
 for the purpose of
either estimating ${\bf X}$ or finding ${\bf K}$ from a
known-plaintext attack. As a result, Bob has a better
channel/observation than Eve. Also, in contrast to classical
cryptography, one can {\em prove} the security of our scheme
against ciphertext-only attacks, although only individual attacks
are described in this paper. One can show that, in a properly
designed system, even an exponentially powerful search with
known-plaintext attacks
 cannot succeed because  Eve does not have ${\bf Y}$ as above. Practically, our scheme can run at high speeds
 because there is no need for  a long  key ${\bf K}$.

 Consider the following scheme in
which each data bit is encoded into a coherent state of an
infinite-dimensional space, referred to as a ``qumode''. As in
\cite{yuenAKE}, there are $M$ possible states \bea \label{mode}
|\alpha_0 (\cos\theta_l +i \sin\theta_l)\rangle, \:\theta_l=2 \pi
l/M \eea for a real $\alpha_0$ and $l\in \{0,\cdots,M-1 \}$,
forming $M/2$ pairs $\{l,l+M/2 \}$. A {\em seed key}  ${\bf K}$ is used
to drive an encryption mechanism whose output is a much longer {\em running
key}  ${\bf K}^{\prime}$ that is used to determine, for each qumode
carrying  bit $b$ $(=0,1)$, which pair of signals (signal set) is to be used.
Each pair may be macroscopically distinguishable since the inner
product of any two basis states is $\exp(-2 |\alpha_0|^2)$. For
large $M$, a lower bound on the obtained mean-square error
$(\delta \theta)^2$ \cite{yuen2} that goes as $1/|\alpha_0|^2$
shows that asymptotically when $M\gg|\alpha_0|$ the attacker's
error probability $P_e^E$ tends to $1/2$, the guessing level, in
an individual attack on the data bit $b$. That this result holds
in the limit $M\rightarrow \infty$ for fixed $|\alpha_0|$ is
intuitively obvious. A two-mode coherent-state realization similar
to Eq. (\ref{mode}), with $|\alpha_0 \cos\theta_l \rangle
|\alpha_0 \sin\theta_l \rangle$, can also be used and the modes
can be interpreted as ones of polarization, time, frequency, or
whatever.

\begin{figure}
\centerline{\scalebox{0.4}{\includegraphics{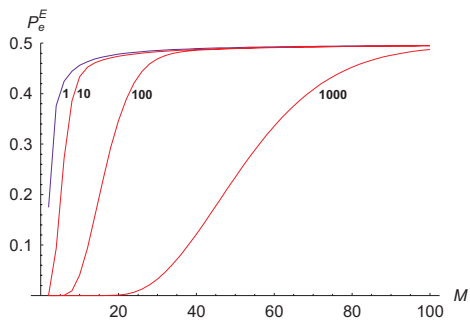}}}
\caption{\label{fig1}
 $P_e^E$ as a function of $M$ for $|\alpha|^2
=1,\: 10,\:100,\:1000$. }
\end{figure}

Numerical calculation of the optimal POVM for individual attack on
bit discrimination for the $M$-ry case has shown \cite{nu-pavia}
that the minimum probability of error $P_e^E$ for an eavesdropper
can be made arbitrarily  close to $1/2$ for a given coherent-state
amplitude $\alpha$. The value $P_e^E\rightarrow 1/2$ for a fixed
average number of photons $|\alpha|^2$ is achieved by increasing
the number of levels $M$. As shown in Figure~\ref{fig1}, $P_e^E$
goes very fast to the asymptotic pure-guessing limit of $1/2$ as
$M$ increases.
The above POVM calculation demonstrates that in this scheme an eavesdropper cannot obtain the bits
sent regardless of the precision of her devices. The optimal POVM gives the maximum amount of
information she could obtain from the sequence of physical signals sent without knowing the key.
This uncertainty is due to the quantum noise of light and cannot be overcome with one's precision
capabilities.
 Bob, on the other hand, by knowing the key  can
extract information with greater precision. His decision has to be
made only between two nearly orthogonal states in the same basis
defined by a given ${\bf K}^{\prime}$. His probability of error is
\cite{helstrom} $ P_e^B= \frac{1}{2}\left(1-\sqrt{1-e^{-2 T
|\alpha|^2}}\right)$, where $T$ is the transmissivity of the
channel. For large values of $|\alpha|$ the minimum probability of
error $P_e^B$ is negligible, which makes possible an excellent
signal recovery by the legitimate receiver. The case of collective
attacks is more complicated and cannot be discussed here, in large
part because there is no meaningful approach for evaluating the
optimal bit error, even in the classical case. However, the
entropy bound (Holevo's theorem) could be used to show the ideal
nature of this scheme for the criterion of data entropy.

The attacker can also try to find the key ${\bf K}$ based on her
copy of the quantum signals, with or without some known-plaintext
(data) corresponding to the signals. Even in a known-plaintext
attack,  the signal quantum fluctuation would yield, from the
number of possibilities in each qumode, an exponential number of
possible ${\bf K}^{\prime}$ in a sequence of data bits. To
identify ${\bf K}$ from such noisy observations of possible ${\bf
K}^{\prime}$ would involve an exponential search, which can always
be launched against a key in known-plaintext attacks but which is
currently believed to be proven impossible computationally even
with a quantum computer.  Note that the attacker has a much more
difficult job of estimating the signal pair from $M/2$ possible
ones, than the user who tries to discriminate 2 possible known
states. A detailed quantitative treatment would be given
 elsewhere.
 It is important to note that, in the case of classical
cryptography, if the running key ${\bf K}^{\prime}$ is used
directly as one-time pad on the data, the result is well known
\cite{menezes} to be insecure against known-plaintext attacks. In
our case, this attack is thwarted by quantum effects as explained
above.

\begin{figure}
\centerline{\scalebox{0.2}{\includegraphics{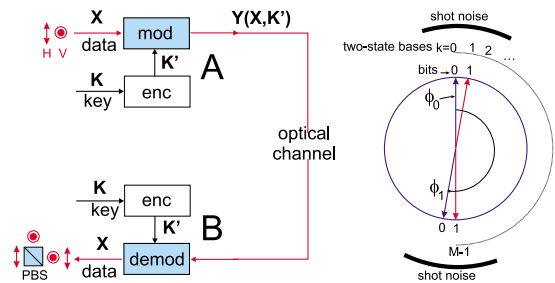}}}
\caption{\label{fig2} {\em Left side:} Basic ciphering scheme with
qumodes of polarization. {\em Right side:} Two neighboring bases
in an M-ry manifold on a great circle of Poincar\'e's sphere. Bits
$0$ and $1$ are antipodals for each basis.    Closest states
(similar polarization ellipses) represent {\em opposite}  bit
values. }
\end{figure}

Consider an implementation of the above scheme as depicted in Fig.
\ref{fig2}. This particular type of KCQ scheme will be refered to
as $\alpha \eta$ (for coherent states and efficiency). In the
encryption scheme for this new protocol, sketched on the left side
in Fig. \ref{fig2},
  Alice uses an explicit short secret key  ${\bf K}$, extended to
a longer key ${\bf K}^\prime$ by another encryption mechanism such as a stream cipher,  to
modulate the parameters of  a multimode coherent state.
For the free-space implementation to be presented, the qumodes are the two orthogonal modes of
polarization. In this case,
 Alice uses the running key ${\bf K}^\prime$ to specify a polarization basis from a set of
$M/2$ uniformly spaced two-mode bases spanning a great circle on
the Poincar\'e sphere, as shown on the right in Fig. \ref{fig2}.
Each basis consists of a polarization state and its antipodal
state at an angle $\pi$ from it, representing the 0 and 1 bit
value for that basis.

The message ${\bf X}$ is encoded as ${\bf Y}({\bf X},{{\bf K}^\prime})$. This mapping of the stream
of bits onto  points on the surface of  the Poincar\'e sphere is the information to be shared by
Alice (A) and Bob (B).  Because of his knowledge of ${\bf K}^\prime$, Bob is able to make a
precise demodulation operation,  producing the plaintext ${\bf X}$. He uses ${\bf K}^\prime$
to apply the requisite polarization transformation to the
 received sequence of polarization states  to return them to the linearly orthogonally polarized
condition, representing the two original bits of the message ${\bf X}$.
Bob's  demodulation is the inverse mapping transformation that was utilized by Alice.

\begin{figure}
\centerline{\scalebox{0.3}{\includegraphics{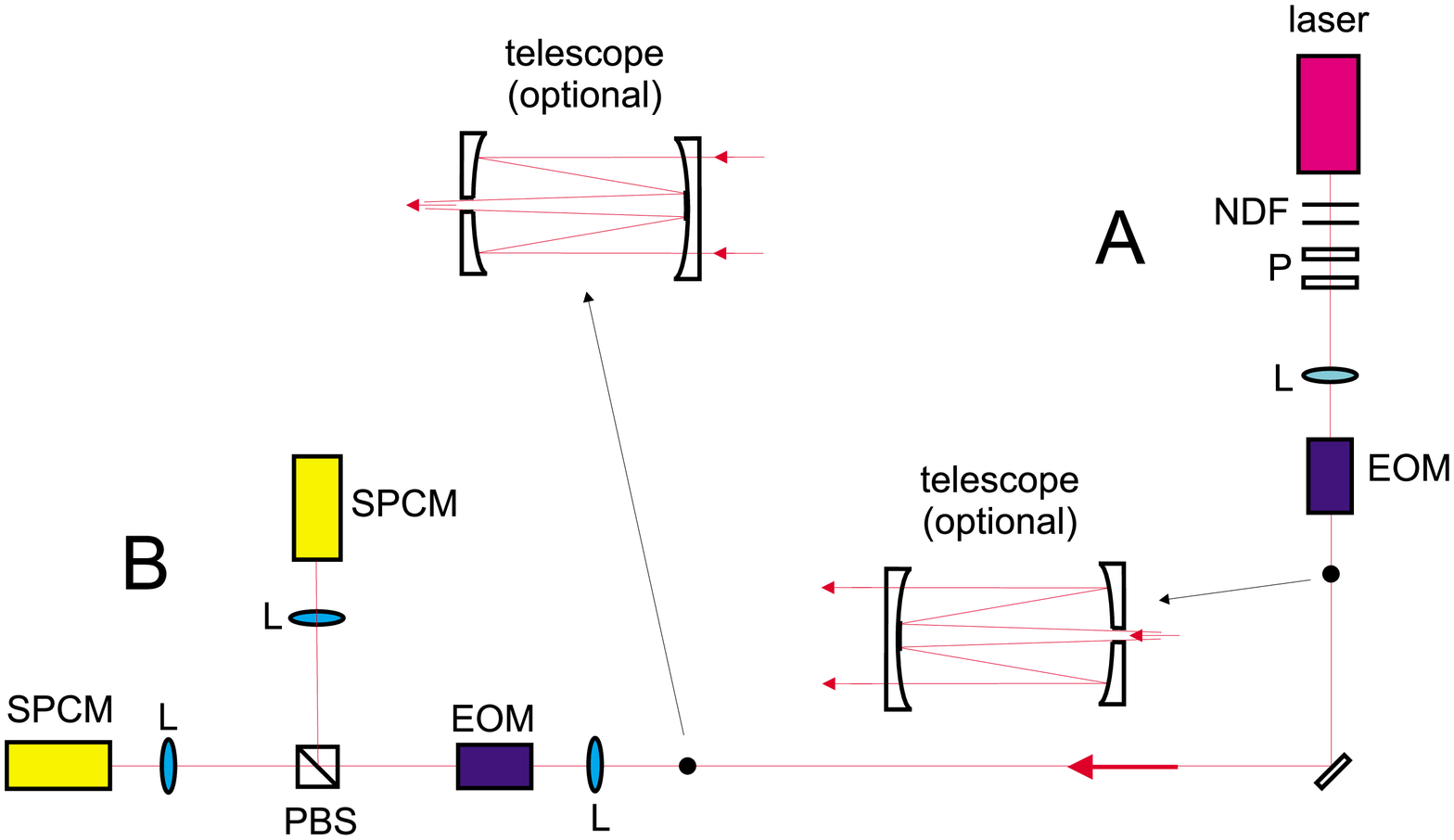}}}
\caption{\label{fig3} Schematic of the experimental setup. SPCM,
single photon counting module; PBS, polarization beamsplitter; L,
lens; P, polarizer; NDF, neutral density filter. Two (optional)
telescopes are shown for field work.}
\end{figure}

Figure \ref{fig3} shows a table-top experimental setup we have
implemented as a proof-of-principle demonstration of this scheme.
The modulation systems utilized both at the transmitter and
receiver ends are electro-optic modulators (EOMs)(New Focus, model
4104), the laser (Toshiba, model TOLD9225M) operates at 670nm,
and the detectors are single photon counting modules
(Perkin-Elmer, model SPCM-AQR-16) with interference filters of
10nm bandwidth in front. A polarization beam splitter (PBS) is
used at the receiver to discriminate between the orthogonal linear
polarization states. Lenses (L) are used to optimize the beam
Rayleigh range within the EOMs. A personal computer containing an
interface card (National Instruments, model PCI-6111E) is used to
control the EOM's (digital-to-analog operation). The same card is
also used for counting the output pulses from the detectors. In
this configuration, a horizontally (H) or vertically (V) polarized
light pulse representing bit 0 or 1 is generated and transformed
into an elliptically polarized light state by application of the
voltage $V_k\:\:(k\in {\bf K}^{\prime})$. This voltage introduces
a phase difference $\Delta \phi_k$ between the physical axes of
the EOM, where $\Delta
\phi_k=\frac{\pi}{2}\frac{V_k}{V_{\pi}}+\phi_0$, and $V_{\pi}$ and
$\phi_0$ are specific to each modulator. This system operates,
with bulk optics,  at 200kHz rate for demonstration purposes and
faster fiber based systems ($\sim$1GHz) are being implemented for
free space as well as fiber channels.

\begin{figure}
\centerline{\scalebox{0.4}{\includegraphics{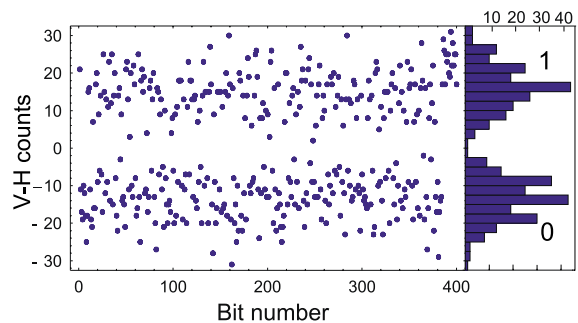}}}
\caption{\label{fig4}  \small Difference of V and H counts (V$-$H)
from Bob's receiver operating at 200kHz, with the average number
of received photons $\langle n \rangle=|\alpha|^2=27$ and $M=50$.
The right inset shows the corresponding histogram indicating clear
separation between the 0 and 1 bit values.}
\end{figure}

\begin{figure}
\centerline{\scalebox{0.4}{\includegraphics{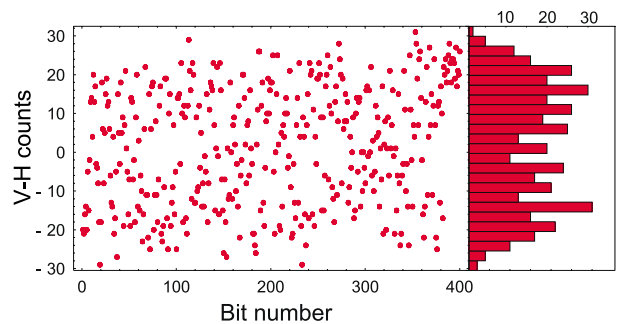}}}
\caption{\label{fig5} \small V$-$H counts from Eve's receiver in
an opaque attack in which she takes all the power from the channel
that would have gone to Bob. All operating parameters are the same
as in Fig. \ref{fig4}, except for Eve lacking the key ${\bf
K}^{\prime}$. The corresponding histogram on the right shows that
distinct bits are not distinguishable by her. }
\end{figure}

Figure \ref{fig4}  shows a sequence of bits as received by  Bob.
The clear separation of the 0 and 1 histograms allow him to make
bit decisions with no error. The same sequence of bits as seen by
Eve are shown in Fig.  \ref{fig5}, giving her a very high
probability of error ($P_e^E\sim1/2$) in bit decisions because of
her lacking the key. $\:$

As an illustration of how  the quantum noise of light can be
utilized by the legitimate receiver on his behalf, in Fig.
\ref{fig6} we show the uncertainty in the polarization angle
produced by a two-mode measurement for an average total photon
number $\langle n \rangle\simeq 38$.
 Regions of low-variance  values are seen around 0 and $\pi/2$ settings of the input polarization
state [cf. Fig. \ref{fig6}(c)]. Such determination of the
polarization angle with angle-dependent uncertainty   shows that a
higher degree of precision in angle determination can be achieved
by an observer with a prior information on how to set the
measuring apparatus than another who  does not have this
knowledge. From Fig. \ref{fig6}(c) one can observe that for
PBS-axis orientations close to the incoming field polarization
direction  the uncertainty in determination of the angle is small.
The polarization angle is obtained by averaging
$\arctan\sqrt{n_V/n_H}$  over the occurrence of $n_H$ and $n_V$,
the photon numbers reaching the detectors [cf. Fig.
\ref{fig6}(b)]. The approximation that $\theta$ can be obtained
through $\arctan\sqrt{n_V/n_H}$ is adequate
 for  mesoscopic signals but  becomes inadequate as $\langle n \rangle \rightarrow 0$
(quantum phase domain).
 Setting PBS without knowledge of this preferred orientation leads, in average, to a larger
error in determination of the polarization angle. On the other
hand, an observer setting his analyzer system close to the field
polarization direction (0 or $\pi/2$)  is only limited by the
optical precision of the analyzer used (which can be made
arbitrarily small), and the noise in the detectors.
\begin{figure}
\label{fig6}
\centerline{\scalebox{0.16}{\includegraphics{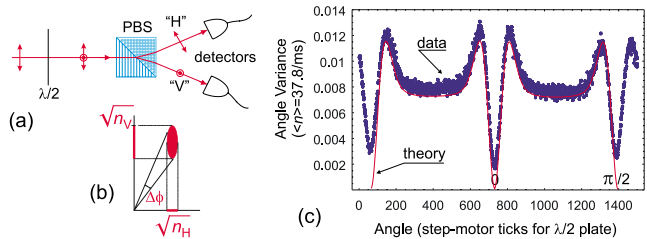}}}
\caption{\label{fig6} \small  (a) Measurement of the polarization
angle through two-mode direct detection with use of a PBS. A
$\lambda/2$ waveplate is rotated to produce different polarization
angles $\phi$.   (b) Uncertainties in $\phi$, $\Delta \phi$,
arise owing to photon-number fluctuations around $\langle
n_H\rangle$ and $\langle n_V\rangle$, where $n_H$ and $n_V$ are
the photon numbers sampled by the two detectors ($\langle n_H n_V
\rangle=\langle n_H\rangle \langle n_V\rangle$). (c) Variance in
the angle $\phi$ obtained via the two-mode measurement versus the
angle set by the $\lambda/2$ plate.   Solid line is the
theoretical prediction, not a fit, {\em without} considering the
detector noises (mainly ``dark'' and Johnson noises), which will
set the ultimate precision limit in these measurements.}
\end{figure}

\begin{figure}
\centerline{\scalebox{0.3}{\includegraphics{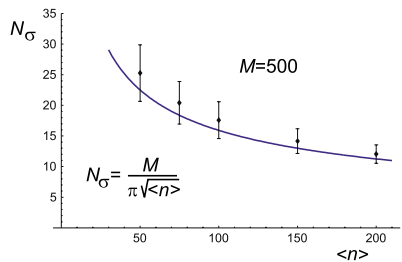}}}
\caption{\label{fig7} Number of bases $N_\sigma$ covered by the
quantum noise as a function of the number of photons $\langle
n\rangle=|\alpha|^2$. For a given $\langle n\rangle$, Alice
repeatedly sets every basis, separated by $\pi/M$ ($M=500$), one
by one, by applying a voltage to the EOM. By measuring $n_H$ and
$n_V$, Eve performs an angle reconstruction in an attempt to
identify the basis used. Eve's standard deviation around the basis
sent by Alice gives $N_\sigma$. Line is the theory.}
\end{figure}

The cryptography system should be designed so that the
uncertainties caused by the quantum noise of light in the
measurement of the polarization angles is large. It can be shown
that the number of bases $N_\sigma$ within a standard deviation of
the measured angle is  $N_\sigma=M/(\pi |\alpha|)$. Fig.
\ref{fig7} shows experimental results that confirm this
dependence.

The effect of noise on signal recovery by an eavesdropper in an opaque attack
can be simulated by sending  repeatedly the same bit,  but varying ${\bf K}^{\prime}$,
from A to B when B (playing Eve) does not apply the key to demodulate the signals.
In the following sequence $S$, the first number in a brace is the basis set by
Alice (selected from  $M=200$ possible positions with $N_\sigma \simeq 6$, and
correctly recovered by Bob with the use of the key) while the second number is
the basis extracted  by Eve through a  {\em single} measurement of $n_H$ and $n_V$.
{\small
$
S= \{ 110,117\} ,\{ 84,78\} ,\{ 108,99\} ,
  \{ 90,91\} ,\{ 100,107\} ,\nonumber
\\ \{ 102,97\} ,
  \{ 84,84\} ,\{ 110,105\} ,\{ 110,111\} ,
  \{ 114,105\} ,\nonumber
\{ 82,86\} ,\\\nonumber
\{ 100,95\} ,
  \{ 92,72\} ,\{ 108,108\} ,\{ 102,90\} ,  \{ 108,97\} ,\{ 96,93\} ,\\\nonumber
\{ 110,103\} ,  \{ 112,121\} ,\{ 86,86\} ,\{ 102,100\} ,\{ 88,91\} ,
  \{ 102,94\} ,\\\nonumber
\{ 106,98\} ,\{ 118,135\} \nonumber.
$
} $\!\!$Clearly, Eve makes a large number of errors in determining
the transmitted bases.

In conclusion, we have demonstrated that under individual attacks
Yuen's encryption protocol is secure  with an adequate number of
bases $M$ and without the need for intrusion detection. Key
expansion is also possible under this scheme, due to the better
observation available to Bob with knowledge of the key.
Significantly, the encryption system allows for signal {\em
amplification}  as long as the security is guaranteed at the
source for the following reason. While Eve has to resolve $M$
levels to tell the bit or ${\bf K}^{\prime}$ without knowing the
key, Bob has to resolve only two levels with the key. Thus,
amplifier noise would hinder Eve's attack while it will not
disrupt Bob's decryption. Furthermore, there is no need to
decrypt/re-encrypt at the nodes of a properly designed
communication line with a moderate number of amplifiers.

{\bf Acknowledgements --}
This work is supported by the DARPA grant F30602-01-2-0528. We thank Adam Rybaltovski
for some help in the early stages of this work and Paul Voss for advice on data
acquisition programming.
$^*$E-mail: barbosa@northwestern.edu.
\vspace{-0.5cm}




\begin{thebibliography}{20}

\bibitem{shannon}
C. E. Shannon,
 Bell Syst. Tech. J. {\bf 28}, pp. 656-715, 1949.

\bibitem{bennett-brassard}
    C. H. Bennett and G. Brassard,
 in Proc. of the  IEEE Int.   Conf. on Computers, Systems and Signal Processing,
Bangalore, India, pp. 175-179, 1984.

\bibitem{alamos}
N. Gisin, G. Ribordy, W. Tittel, and H. Zbinden, Rev. Mod. Phys. {\bf 74}, pp. 145-195 (2002).

\bibitem{schneier}
B. Schneier, ``Secrets and Lies'', Wiley, New York, 2000.

\bibitem{yuenAKE}
H. P. Yuen,
 in Quantum Communications, Computations, and Measurements III,
 Plenum Press, 2001.

\bibitem{yuen_tobe}
H. P. Yuen, to be published.

\bibitem{yuen2}
H. P. Yuen,
 in Quantum Squeezing,
 Springer-Verlag, to be published; also in quant-ph0109054.

\bibitem{nu-pavia}
G. A. Barbosa, E. Corndorf, P. Kumar, H. P. Yuen,
G. Mauro D'Ariano,  M. G. A. Paris, and P. Perinotti, Int. Conf. on Quantum
Communication, Measurement and Computing, July 2002 (Rinton Press).

\bibitem{helstrom} C. W. Helstrom, {\em Quantum Detection and Estimation Theory},
(Academic Press, 1976).

\bibitem{menezes}
A. J. Menezes, P. C. van Oorschot, and S. A. Vanstone, ``Handbook of Applied Cryptography'',
 CRC Press, New York, ch. 6, 1997.





\end{thebibliography}
\end{document}